\documentclass[conference]{IEEEtran}

\usepackage{cite}
\usepackage{amsmath,amssymb,amsfonts}
\usepackage{algorithmic}
\usepackage{graphicx}
\usepackage{textcomp}
\usepackage{url}
\usepackage{xcolor}
\usepackage{booktabs}
\usepackage[ruled,linesnumbered,nofillcomment]{algorithm2e}

\SetCommentSty{mycommfont}
\usepackage{multirow}
\def\BibTeX{{\rm B\kern-.05em{\sc i\kern-.025em b}\kern-.08em
    T\kern-.1667em\lower.7ex\hbox{E}\kern-.125emX}}
\usepackage{caption}
\begin{document}

\title{On the Efficiency of Decentralized File Storage for Personal Information Management Systems\footnotemark}

\author{
\IEEEauthorblockN{Mirko Zichichi\IEEEauthorrefmark{1}\IEEEauthorrefmark{2}\IEEEauthorrefmark{4}, Stefano Ferretti\IEEEauthorrefmark{3}, Gabriele D'Angelo\IEEEauthorrefmark{2}}
\IEEEauthorblockA{\IEEEauthorrefmark{2}Department of Computer Science and Engineering, University of Bologna, Italy\\ \emph{g.dangelo@unibo.it}}
\IEEEauthorblockA{\IEEEauthorrefmark{3}Department of Pure and Applied Sciences, University of Urbino "Carlo Bo", Italy\\ \emph{ stefano.ferretti@uniurb.it}}
\IEEEauthorblockA{\IEEEauthorrefmark{4}Ontology Engineering Group, Universidad Politécnica de Madrid, Spain\\ \emph{mirko.zichichi@upm.es}}
}

\maketitle 
\let\thefootnote\relax\footnotetext{\IEEEauthorrefmark{1}This work has received funding from the European Union’s Horizon 2020 research and innovation programme under the Marie Skłodowska-Curie ITN EJD grant agreement No 814177 Law, Science and Technology Joint Doctorate - Rights of Internet of Everything }

\footnotetext{
\textbf{{\color{red} This is the pre-peer reviewed version of the article: ``Mirko Zichichi, Stefano Ferretti, Gabriele D'Angelo. On the Efficiency of Decentralized File Storage for Personal Information Management Systems. Proceedings of the 25th IEEE Symposium on Computers and Communications (ISCC 2020).''.}}}

\IEEEpubidadjcol

\begin{abstract}
This paper presents an architecture, based on Distributed Ledger Technologies (DLTs) and Decentralized File Storage (DFS) systems, to support the use of Personal Information Management Systems (PIMS). DLT and DFS are used to manage data sensed by mobile users equipped with devices with sensing capability. DLTs guarantee the immutability, traceability and verifiability of references to personal data, that are stored in DFS. In fact, the inclusion of data digests in the DLT makes it possible to obtain an unalterable reference and a tamper-proof log, while remaining compliant with the regulations on personal data, i.e. GDPR. We provide an experimental evaluation on the feasibility of the use of DFS. Three different scenarios have been studied: i) a proprietary IPFS approach with a dedicated node interfacing with the data producers, ii) a public IPFS service and iii) Sia Skynet. Results show that through proper configuration of the system infrastructure, it is viable to build a decentralized Personal Data Storage (PDS).
\end{abstract}


\section{Introduction}

The advent of social media and Web 2.0 favoured a process that broke the boundaries between authorship and readership: users produce the data that is consumed by other users. This has increased the privacy threats of applications that are shaped by user-generated content, as it often consists of highly personal data.
In general, the economics of personal information is helped by the more pervasive nature of today's digital world. This information enables organizations to provide personalized or more useful services in digital and physical spaces, but it could also have potentially harmful consequences for the privacy and autonomy of users and society at large.
Current platform-centered data management techniques threaten the control that individuals exercise over their personal information and give to few companies the power to necessarily rely on them to explore, filter and obtain data of interest. Not mentioning the fact that some of these central entities can operate without any transparency on the use of users' data.

An individual digital counterpart can be depicted not only by using his own personal information, but also that of his social links (e.g. friends, family, colleagues) . Thus, it becomes easier to understand users activity choice and lifestyle patterns \cite{hasan2016understanding} and to make more intrusive recommendations using this data \cite{partridge2009enhancing,bothorel2018location}.
Lack of privacy control, for instance, leads an individual being thrown into a ``filter bubble'' that can affect his ability to choose how he wants to live, simply because the platforms that build this bubble choose which options he can be aware of \cite{pariser2011filter}. On a social level, this scheme can lead to a deeper polarization and manipulation of society \cite{cadwalladr2018revealed, christl2017companies}.

On the other hand, Internet of People (IoP) is emerging as a paradigm that will leverage such centralized platforms, when needed, and will work on top of the Internet to place individuals and their personal devices at the heart of the data management design \cite{conti2018internet}. Smartphones and personal IoT devices will function as gateways, being the proxies of their users in the digital world.
Thus, user devices play a more active role on the data management, without delegating the whole management process to centralized remote platforms. 

Crowd-sensed information is essential for building sophisticated smart services raising awareness about the environment, e.g.~to improve Intelligent Transportation Systems (ITS) \cite{ieeeaccess2020}. We thus envisage that users will be able to record data, store them in some Personal Data Storage (PDS), and communicate with other users as well.
By promoting a distributed storage and computing network, the ability of these centralized strongholds to use users' personal information in the digital advertising industry could be transferred to those users that are directly concerned.
As mentioned, we are dealing with services where data generated by users' smartphones, vehicles' sensors or IoT devices, are transformed into new meaningful information useful for individuals and the ecosystem itself \cite{PrivacyMONET2020}. Hence, one of the main issues is to provide means to easily publish data, while granting compliance with the (related) individuals' privacy preferences and regulations, i.e.~the GDPR \cite{eu-679-2016}.
A Personal Information Management System \cite{edps2016} model can provide users with tools for managing the collected data and access control to other parties wishing to use such data, as well as supporting incentives for all stakeholders. 
The PIMS adheres to the rules on the transmission and processing of personal data brought by the GDPR, acting as a strong facilitator for the consent of individuals, required for purposes of direct marketing, behavioral advertising, location-based advertising or digital market research based on tracking.
 
Decentralized architectures might promote individuals' data sovereignty and the possibility of the creation of fair data marketplaces, where individuals share their data and access data, as data consumers, with permissions granted following an agreement \cite{dltancona}.
In this paper, we describe a decentralized software architecture for a PDS based data sharing and trading, whose main building blocks are Distributed Ledger Technologies (DLTs) and Decentralized File Storage (DFS). Data sharing services are defined to let users share their data. These services permit to define how data can be shared, but also how data are acquired. Access to crowd-sensed data is regulated through smart contracts, that implement a control list and provide access only to authorized users.

We provide experimental results of a real testbed evaluation of the critical aspect of the use of DFS to store user generated data.
In particular, through a trace-driven simulation, we instantiated an ITS application.
We generated a data traffic mimicking users traveling in public transport in Rio de Janerio, that periodically sense data and send them to their PDS. Such data traffic was submitted to the IOTA DLT and the employed DFS, under different real setups. Outcomes demonstrate that, through proper configuration of the system infrastructure, it is viable to build a decentralized Personal Information Management System (PIMS).

The remainder of this paper is organized as follows. Section~\ref{sec:back} introduces some background and related work. Section~\ref{sec:archi} outlines the reference distributed software architecture. Section~\ref{sec:eval} describes the design of the experimental evaluation we conducted and the obtained results. Finally, Section~\ref{sec:conc} provides some concluding remarks.

\section{Background and Related Work}\label{sec:back}

\subsection{Distributed Ledger Technologies}
A Distributed Ledger Technology (DLT) is a technical implementation of a data ledger, thought with the aim to move trust from a human intermediary, that manages a transaction between two parties, to a protocol that allows the two parties to transact directly, i.e.~without the need of a third party \cite{ccnc2020}. The ledger ensures immutable persistence of data, thus providing untampered data to applications when it is necessary. For this reason, DLTs represent an appealing technology for the development of trustful and reliable decentralized Personal Information Management (PIM) services \cite{ieeeaccess2020,vanderHeijden:2017}.

\subsubsection{Smart Contracts}
Smart contracts provide a new paradigm where an immutable set of instructions is deterministically executed during a transaction between two parts. Without the presence of a third party, the execution of a smart contract is performed in such a way that a contract issuer can always be sure that the behavior he implemented is observed. In the case of Ethereum~\cite{buterin2013ethereum}, every process is completely traced and permanently stored in the blockchain. Moreover, the smart-contract computation is executed by all network participants. Ethereum provides a distributed virtual machine able to process any kind of computation through smart contracts. However, it is well known such blockchain has some scalability issues \cite{bez2019scalability}. Conversely, the IOTA ledger~\cite{popov2016tangle} is thought to provide better scalability, but it currently does not support smart contracts.

\subsubsection{IOTA}
IOTA is a permissionless DLT that allows hosts in a network to transfer immutable data among each other. It is specifically designed for the IoT industry. The ledger used in IOTA is not structured as a blockchain but as a Direct Acyclical Graph (DAG) called the Tangle~\cite{popov2016tangle}. In the IOTA DAG, the graph vertices represent transactions and edges represent approvals. When a new transaction is issued, it must approve the two previous transactions and the result is represented by means of directed edges.
The validation approach is thought to address two major issues of traditional blockchain-based DLTs, i.e.~latency and fees. IOTA has been designed to offer fast validation, and no fees are required to add a transaction to the Tangle \cite{BROGAN2018257}.

An important feature offered by IOTA is the Masked Authenticated Messaging (MAM), a communication protocol that adds the functionality to emit and access an encrypted data stream over the Tangle. Data streams assume the form of channels, i.e.~a linked list of ordered transactions. Once a channel is created, only the channel owner can publish encrypted messages on it. Users that possess the MAM channel encryption key are enabled to decode the message. MAM enables users to subscribe and follow a stream of data, generated by some device. 

\subsection{Decentralized File Storage (DFS)}
Decentralized file storage is a potential solution for maintaining files in a system without having to rely on a large, centralized silos that may not completely assure privacy of information. Such technologies are fundamental for DLTs, since these can be used when the ledger and the consensus mechanism disincentives data storing.

\subsubsection{IPFS}
The InterPlanetary File System (IPFS)~\cite{benet2014ipfs} is a protocol that builds a distributed file system over a peer-to-peer network. IPFS creates a resilient file storage and sharing system, with no single point of failure and without requiring mutual trust between nodes. This technology is useful to store data that is not convenient to put on DLTs. Files published in the IPFS network take the form of IPFS objects. In order to retrieve an object, only the file digest is needed, i.e.~the result of an hash function applied on the file. Put in other words, the file digest is the identifier of the IPFS object. Users that want to locate that object use this identifier as an handle.

\subsubsection{Sia}
In order to provide incentives to nodes for maintaining data, some DFSes integrate DLTs, bringing together clients' requests with storage nodes' offers. 
For instance, Sia \cite{vorick2014sia} is a DFS that integrates a blockchain in order to reward hosts for keeping files. It uses File Contracts, i.e.~a particular kind of smart contract employed to arrange an agreement between a storage provider and their clients. Sia is very promising but, at the time of writing, it lacks the simplicity and the level of maturity provided by IPFS. Probably for this reason, current solutions in literature are mostly based on IPFS~\cite{naz2019secure,hawig2019designing,8400511}.

\subsection{Personal Information Management Systems and GDPR}
Personal Information Management Systems (PIMS) \cite{edps2016}, built on top of the IoP paradigm, can serve as a unique digital space fully managed by individuals that exploit PDS. It is a model that provides means for individuals to reflect on their online presence, restore sovereignty over their data, and enable a process of negotiation with other parties concerning personal data. This is also known as databox \cite{perera2017valorising,Crabtree2018}.
The PIMS model is largely symbolic at the moment, but it is not a theoretical model. 
Solid \cite{sambra2016solid} is a related implementation, born with the purpose of giving users tools for letting them choose where their data resides and who is allowed to access and reuse it. Solid involves the use of distributed technologies and Semantic Web integration to store data in an online storage space called Pod.

Some DLT features come into conflict with GDPR compliance, therefore special consideration must be given when developing new designs. Onik et al.~\cite{onik2019privacy} propose a model that traces the life cycle of personal data that are stored ``off-chain", i.e.~not directly stored in the DLT, in order to respect the right to be forgotten prescribed by the GDPR. In their architecture, smart contracts contain the terms and consent for the use of personal data of the data subjects, i.e.~the individuals who are identified by such data, which must be accepted by the data processors.
Ahmed et al.~\cite{ahmed2020towards} focus on online social networks and their lack of GDPR compliant consent management mechanisms. They present some opportunities for using DLT to address this issue and to provide fine-grained control over personal data, while discussing also the challenges of DLTs under GDPR.

\section{System Architecture}\label{sec:archi}

\begin{figure}[t]
  \centering
  \includegraphics[width=\linewidth]{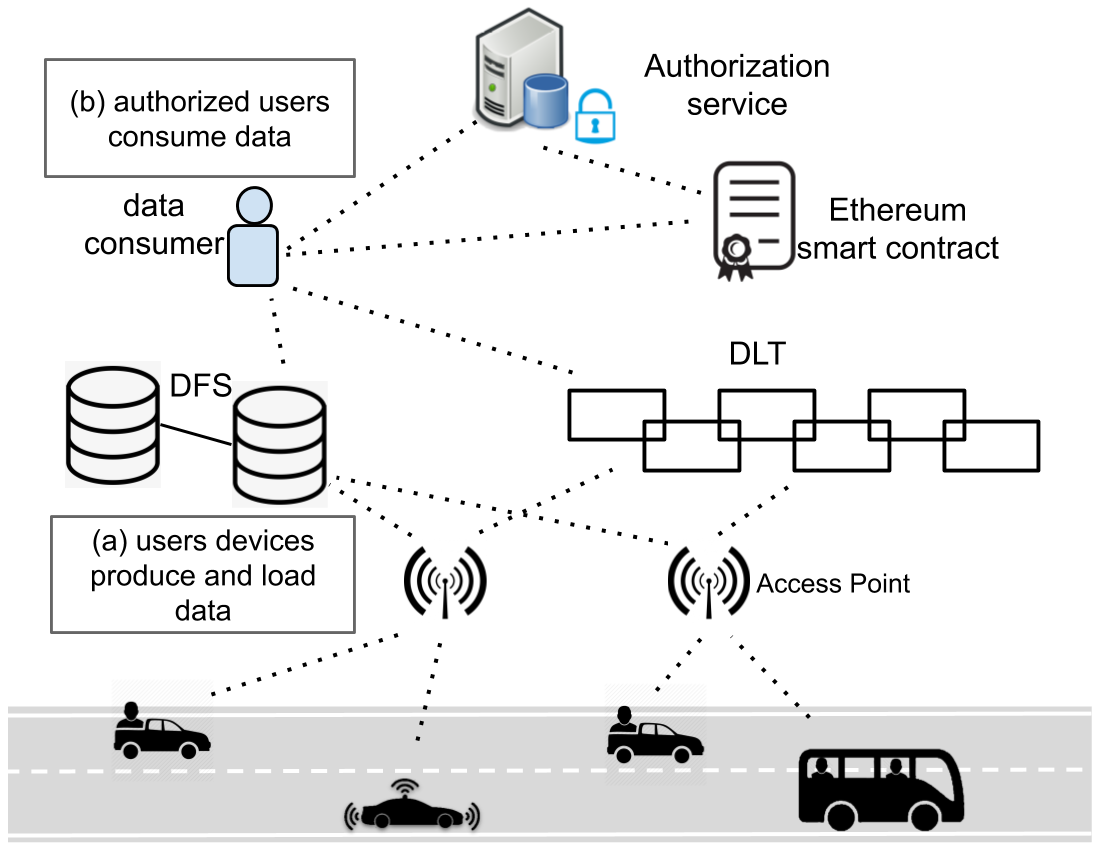}
  \caption{System architecture.}
  \label{fig:archi}
\end{figure}

We consider a system where mobile users generate data through their devices sensors and store it in a PDS. Figure \ref{fig:archi} (lower part) shows an example of mobile users that generate data, while moving in some vehicles, and issue such data into the system (in the figure, we specifically focus on vehicles, due to the particular simulation we used in the performance evaluation). 
The aggregation of different distributed technologies enables the collection and sharing of crowd-sensed and user-managed data, which are used to build services and applications according to the IoP paradigm.
Providers of such services can use the data submitted by users to gain knowledge of a particular area and develop geolocalized smart services built using smart contracts as business logic.

Crowdsensed data are stored in a DFS and then referenced in a DLT (in the middle of Figure \ref{fig:archi}). 
Storing data into a DFS usually requires lower latencies with respect to those that can be obtained using DLTs. (In fact, DLTs typically require some time consuming Proof-of-Work.) 
However, validation is obtained through the publication of the data digest into a DLT.
Personal data must not be stored directly in the DLT, even when encrypted, because of the right to be forgotten brought by the GDPR.
Therefore, in general, we only consider DFS for data storage, but we adopt the following heuristics to quicken the process of publishing non-personal data:
\begin{itemize}
    \item Personal data and large sized non-personal data is stored into a DFS and referenced in the DLT through its digest.
    \item Small sized non-personal data (whose size is comparable to the size of its digest) is directly stored in the DLTs.
\end{itemize}

Once a file is published in the DFS, the returned reference can be employed to retrieve it.
Taking, for instance, IPFS \cite{benet2014ipfs} as the used DFS, such reference is, in fact, the data digest itself that it is stored in the DLT. Thus, the piece of data is published as an IPFS object and then (asynchronously) referenced through its hash into a MAM transaction. The digest allows verifying the integrity of the IPFS object. To upload files on IPFS, a node running the IPFS protocol is necessary. Due to the fact that it is (still) not feasible to run an IPFS node on constrained devices (such as smartphones or sensors), other solutions must be explored. For example, in our architecture we assume that an IPFS service provider (e.g.~Infura \cite{infura}) lets a user permanently store files in the IPFS network, as long as they reach an agreement (e.g.~by paying a subscription). 

\subsection{Data Validation Through DLTs}
DLTs allow avoiding all the typical drawbacks of server-based approaches (e.g.~censorship, single point of failure), and offer features such as data immutability, verifiability and, most importantly, traceability. These can be used to obtain immutable references to users' data and provide a tamper-proof log, which can be consulted in case of a dispute. 

During the implementation of the system architecture, we decided to employ IOTA as DLT. 
IOTA exploits the Tangle as the public data ledger, accessible by anyone.
The Tangle stores immutable information that cannot be censored/removed.
MAM addresses the prime requirement of data protection, as MAM channels are used to store data using encryption and providing access only to eligible users.

\subsection{Data Access}
The use of data is authorized only to entitled users (upper part of Figure \ref{fig:archi}). Access control is performed through smart contracts \cite{ieeeaccess2020}. Access to the data can be purchased or can be allowed by the owner through dedicated smart contract methods. If access is purchased, in the Ethereum, such methods take the form of payable functions that enact monetary transactions. Hence, due to the presence of smart contracts, no direct interactions are needed among the data owner and users interested in his data.
In practice, each kind of data (MAM) channel in IOTA is associated to a specific smart contract in Ethereum.
The smart contract maintains an Access Control List (ACL) that represents the rights to access a bundle of data. This bundle is composed of references to MAM channels or to single channels messages.

Once a consumer is eligible to obtain certain data, i.e.~he is in the ACL, he can access such data through an access key, which is provided by an authorization service. 
The consumer sends a request to the authorization service. Upon request, the authorization service checks if he is eligible, through interaction with the smart contract. If this is the case, the authorization service provides the user with the related access keys.
In particular, for each MAM message and its related DFS object (if available), there is a key, which is used to encrypt the produced data.

In this paper, we will not go into the details of the authorization service, since the main focus is on the performance of the DFS component. However, there are several methods to design an authorization service. The simplest solution can resort to a a traditional Client/Server approach, where a server provides the authorization service and holds the entire set of secret keys to access MAM messages and IPFS objects. 
However, more sophisticated methods can be considered, based for instance on proxy re-encryption or secret sharing \cite{ieeeaccess2020}.

\section{Performance Evaluation}\label{sec:eval}

The main critical points of the devised software architecture, that need to be studied to assess its scalability, are concerned with the responsiveness and reliability of both DTLs and DFS systems. While we already studied the behavior of DLTs in \cite{dlts}, in this work we focus on DFS.

\begin{figure}[t]
  \centering
  \includegraphics[width=.90\linewidth]{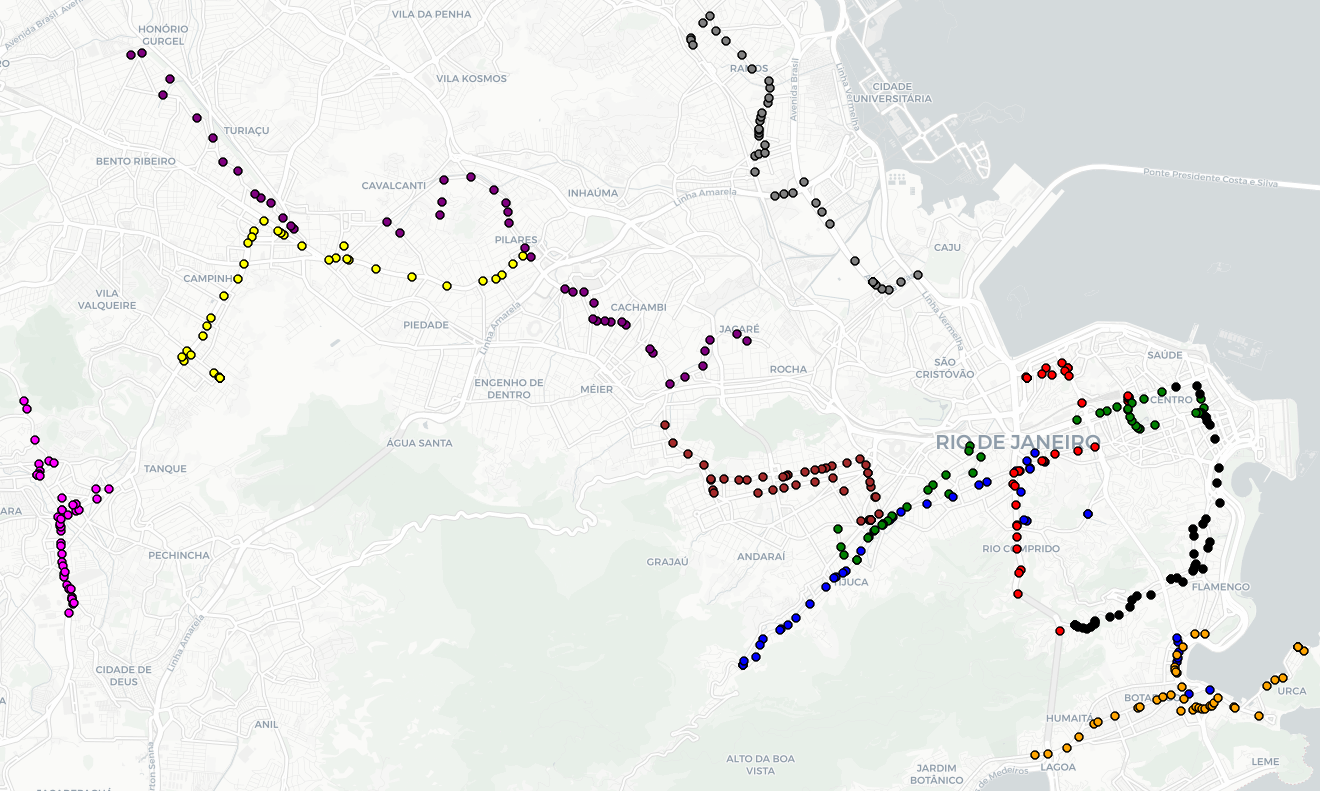}
  \caption{The 1 hour long path of 10 buses in Rio de Janeiro (Brazil).}
  \label{fig:buses}
\end{figure}

\begin{figure*}[t]
  \centering
  \includegraphics[width=.85\linewidth]{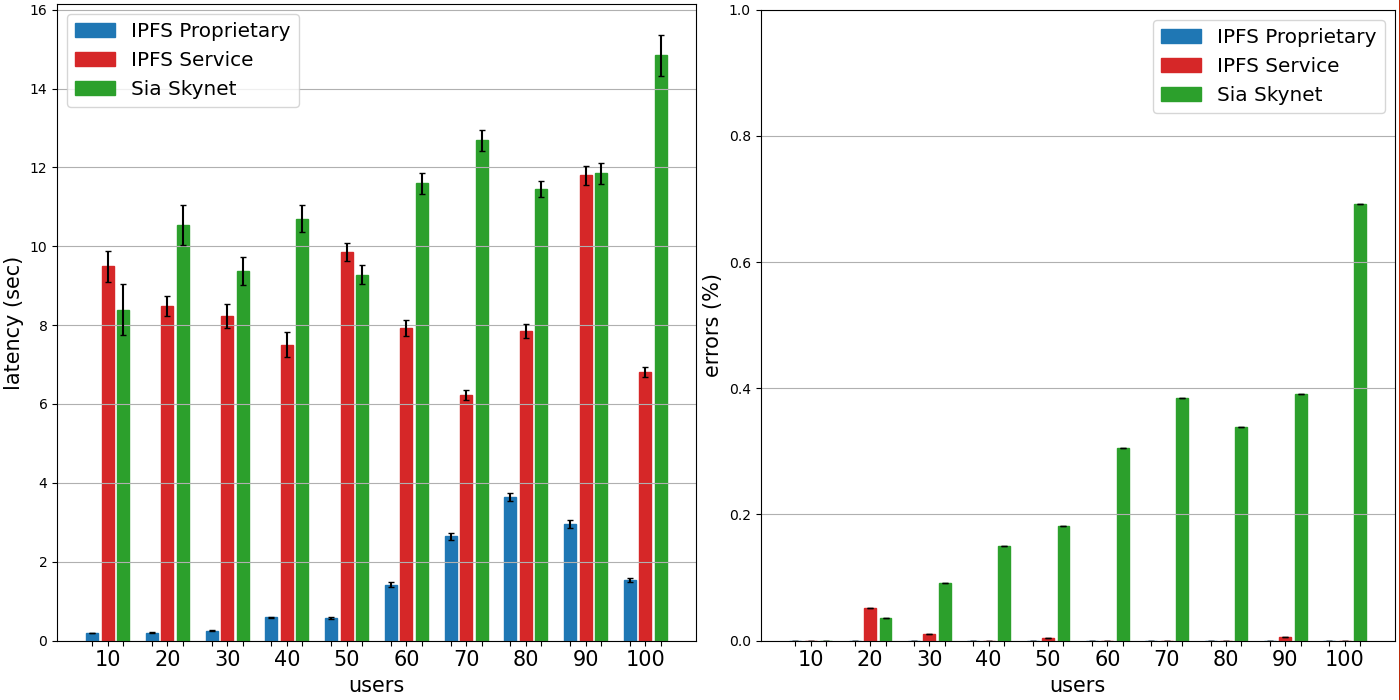}
  \caption{Latencies and errors when sending messages to DFS nodes. Black line represents the confidence interval (95\%) }
  \label{fig:smalllat}
\end{figure*}
\begin{figure*}[th]
  \centering
  \includegraphics[width=.85\linewidth]{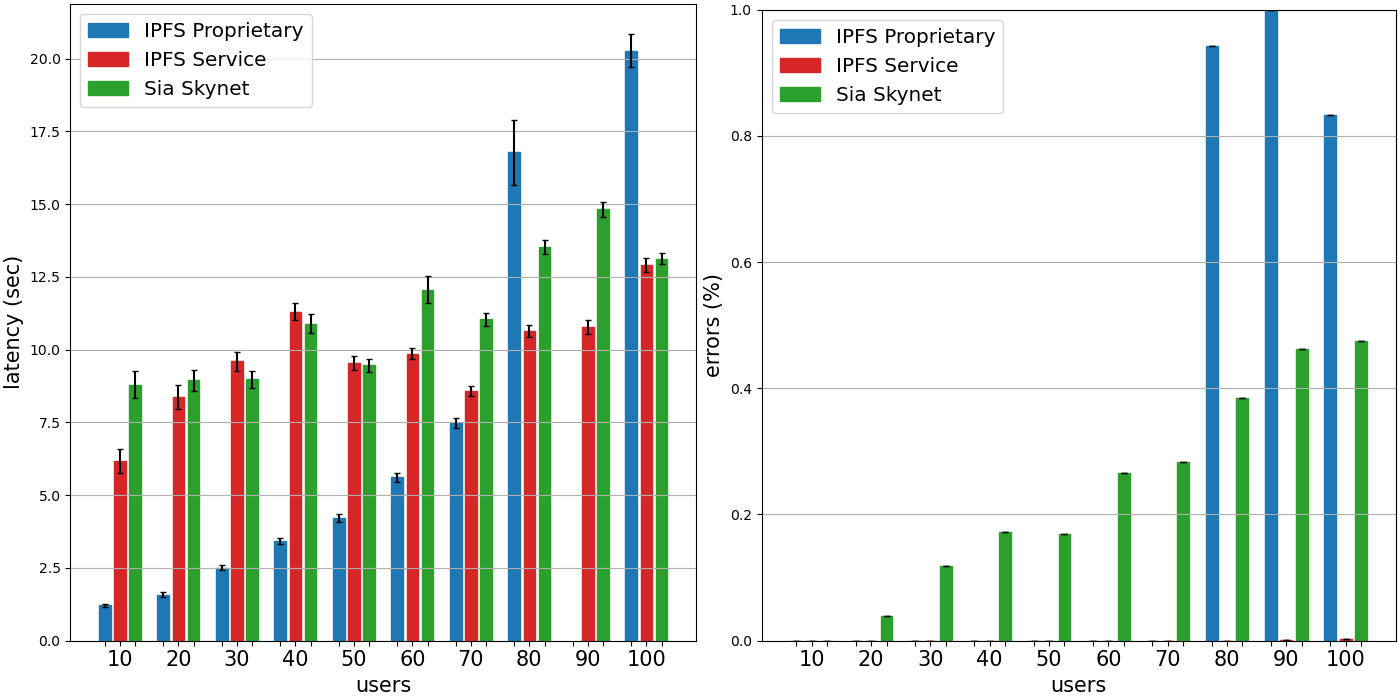}
  \caption{Latencies and errors when sending photos (1 MB) to DFS nodes. Black line represents the confidence interval (95\%) }
  \label{fig:biglat}
\end{figure*}

Our experimental scenario was based on a hypothetical real ITS application. In particular, we conducted a trace-driven experimental evaluation.
Traces were generated using the RioBuses dataset, a real dataset of mobility traces of buses in Rio de Janeiro (Brasil) \cite{coppe-ufrj-RioBuses-20180319}. Based on these traces, we simulated a number of users' devices on board of buses that, during their path, periodically generate sensed data. We considered one user for each bus.
These data may represent temperatures, air pollution values, etc. In this case, we focused on two different types of data:
\begin{itemize}
    \item \textbf{Small sized data}: such as geolocation, i.e. latitude and longitude (100 bytes), encoded as a JSON. 
    \item \textbf{Large sized data}: photos (1 MB).
\end{itemize}

Figure \ref{fig:buses} shows the paths of 10 buses, as an example, that were considered during our tests.
Messages storing geolocation data or a photo sensed by the users' devices were utilized to generate real requests transmitted to the DFS.

In this assessment, we used a single DFS node, while varying the number of users,~i.e. we tested different cases with a specific amount of users associated to a single DFS node. We compared two DFSes solutions: IPFS and Sia \cite{vorick2014sia}. The idea is to evaluate the available solutions to store data in a DFS, comparing the latencies to request IPFS nodes and Sia nodes.
In the case of IPFS, we assessed two different scenarios: i) the case with a dedicated IPFS node, devoted to handle only requests coming from our application (referred as ``IPFS Proprietary''), ii) the case with a public IPFS node, that can be contacted also by other applications (referred as ``IPFS Service''). In the case of Sia, we exploited the Skynet platform services 
to easily access the provided permanent storage. In particular:
\begin{itemize}
    \item \textbf{IPFS Proprietary}: We setup an IPFS node on a dedicated device (dual core CPU, 8GB RAM), connected to other nodes in the main network. Thus, the host simulating the users' devices was the only one sending requests for storing files on it.
    The files are stored locally (and on its IPFS neighbors). Each file is maintained as long a node is incentivized to ``pin'' it, i.e.~to keep it, it remains available to anyone.
    \item \textbf{IPFS Service}: 
    We tested the generic Infura service provider \cite{infura}, that offers a free access to IPFS.
    \item \textbf{Sia Skynet}: 
    Tests are conducted making requests to a node in the Skynet, a content delivery platform built on top of Sia. A Skynet node is a special Sia node that has already formed contracts with every available host, paying for the files uploaded to them, and thus proposing a service with its own policies on how many and what types of files you can upload 
\end{itemize}

The methodology followed to run the tests is the following: tests were conducted in order of dimension (small files first, then larger ones) and users number (from 10 to 100). The performance evaluation has been designed as stress test in which each simulation sends requests to the three different types of DFS nodes following the buses real traces. A simulation lasts 15 minutes and sends exactly 15 messages for each user. An interval of 10 or 20 minutes has been applied to separate consecutive runs of the simulation. The complete dataset and the scripts being used are stored in a github repository \cite{githubrepo}.

\subsection{Results}
Figure \ref{fig:smalllat} shows the latency obtained when sending small messages to the considered DFS nodes and the related percentage of errors. In case of error, the node almost always responds with a HTTP status codes such as 500 or 504. (In this case, the message is not considered when averaging latencies.)
Similarly, Figure \ref{fig:biglat} shows the latencies and errors when larger files, i.e., a 1 MB-sized photo, are sent to the DFS nodes. Each histogram bar shows the average value obtained in the specific configuration with its confidence interval (95\%).

In general, we noticed better performance for the IPFS technology, especially when a dedicated node is employed. In fact, IPFS Proprietary has an average latency of about $1$ sec in the case of small data, with a very limited confidence interval. Conversely, both IPFS Service and Sia Service do have higher latencies and confidence interval. As far as errors are concerned, we noticed a low level of errors in IPFS configurations and a high error rate with Sia.

The results show that IPFS Service and Sia have a similar behavior with both small and large files. This is due to the fact that both have more resources than the proprietary node, which is then unable to cope with larger files. More specifically, the IPFS Service has similar or better performance than Sia, with a very low error rate. On the other hand, in order to maintain a stable latency in responses the Sia node shows an increase in errors that seems to grow linearly with the number of users.

In the stress test that we implemented, the IPFS Proprietary performances get worse when increasing the number of users. Since our tests are performed in sequence by varying the number of users from 10 to 100, what happens is that the node running IPFS accumulates the workload from previous tests, resulting in a cascading effect on the following test. Since service nodes (i.e.~IPFS Service and Sia Service) have more resources this effect is less evident, but for the proprietary node there is a turning point with 80 users where, overall, performance degrades in the presence of large files, while latencies with small files remain stable (or even decrease). In the case of large files this behavior even leads to a complete failure of all requests in the 90 users tests. Presumably, the request rate produced between the 70 and 80 users simulations (obtained from the real traces) contains an accumulation of workload that cannot be dealt with within a short time. In that case, the IPFS node tries to distribute the files over the network, but at the same time receives too many requests.

In general, IPFS Proprietary always works better except for over 80 users in the case of large files. This means that a dedicated node is always preferable, but must be limited to a rate of 60-70 users requests per minute. 

This suggests that, in presence of a properly tuned infrastructure based on edge computing, with a proper deployment of proprietary DFS nodes devoted to handle the communications with a controlled set of users, then PIMS can be adequately supported by the infrastructure.

\section{Conclusions}\label{sec:conc}

In this paper, we have presented a general architecture based on DLTs and DFS for the development of a decentralized Personal Information Management System (PIMS). We specifically focused on the issue concerned with Personal Data Storage (PDS) services. 
We considered a specific use case, related to Intelligent Transportation Systems (ITS), in order to perform a real performance evaluation of three different DFS approaches. More specifically, we contrasted two online services, i.e.~Infura IPFS, Sia Skynet, and a proprietary service where a dedicated node was in charge of running the IPFS and offering access to the peer-to-peer infrastructure.
Results show that the three approaches provide different performances. In particular, up to a certain overload, the proprietary solution seems to offer better guarantees in terms of responsiveness and reliability. 
This suggests that through a proper configuration of the system infrastructure, it is possible to build scalable and reliable decentralized systems for personal information management, that at the same time guarantee data sovereignty.

\bibliographystyle{IEEEtran}
\bibliography{biblio}

\end{document}